\documentclass[preprint2]{aastex631}

\usepackage{graphicx} 
\usepackage{natbib}
\usepackage{amsmath}
\usepackage{hyperref}
\usepackage{gensymb}
\usepackage{hyperref} 
\usepackage{enumitem} 
\usepackage{mathtools}
\usepackage{xcolor}

\newcommand{\kms}{km~s$^{-1}$}
\hypersetup{
  citecolor=blue, 
}
\begin{document}
\newcommand{\red}[1]{\textcolor{red}{#1}}

\title{Structure and dynamics of erupting solar prominences using the Rolling Hough Transform: \\ Toward a feature-oriented classification}

\correspondingauthor{Harry Birch}
\email{harry.birch@northumbria.ac.uk}

\author[0009-0000-4625-1735]{Harry Birch}
\affiliation{Department of Mathematics, Physics and Electrical Engineering, Northumbria University, Newcastle upon Tyne, NE1 8ST, UK}

\author[0000-0001-8954-4183]{St\'ephane R\'egnier}
\affiliation{Department of Mathematics, Physics and Electrical Engineering, Northumbria University, Newcastle upon Tyne, NE1 8ST, UK}

\begin{abstract}

The classification of solar prominences has proven to be challenging due to their diverse morphologies and dynamical behaviour. Complexity is heightened when considering eruptive prominences, where the dynamics demand methods capable of capturing detailed structural information. While there exists a range of line-of-sight (LOS) and plane-of-sky (POS) techniques which have advanced our understanding of prominence motions, they are subject to limitations, emphasising the need for effective methods of extracting structural information from prominence dynamics. We present a proof-of-concept for the spatial Rolling Hough Transform (RHT) algorithm, which identifies fine-scale structural orientation in the POS, applied to prominence structure and dynamics. We demonstrate the RHT approach using two contrasting prominence dynamics events using SDO/AIA 304~\AA\ observations: (1) a quiet-Sun eruption, (2) activation (swirl) of a polar-crown prominence. By analysing the light curves and movies from each event, we divide the events into distinct dynamical phases: from slow rise to drainage. The spatial RHT method enables us to extract structural information and localised dynamics for both events and the different evolution phases. We develop a classification to label the prominences as either radially or tangentially oriented structures. The quiet-Sun eruption has a predominately tangential structure in the slow-rise phase, but displays greater radial features during/after the eruption. The polar-swirl activation initially shows a strong radial contribution, which diminishes as more tangential structures appear during/after the activation. Our results demonstrate the successful application of the spatial RHT to prominences, leading to the classification of individual prominences and an insight into their dynamics.

\end{abstract}



\section{Introduction} \label{sec:Sect1} 

Solar prominences are features of photospheric/chromospheric plasma material at different ionisation degree, found suspended in the solar corona \citep{2014LRSP...11....1P,2010SSRv..151..243L,2010SSRv..151..333M}. They are subject to a broad range dynamics, both: (i) small-scale dynamics within the prominence such as MHD waves, coronal rain, nanojets, (ii) bulk dynamics often leading to prominence eruptions (PEs). Such eruptions are closely related to coronal mass ejections and flare activity \citep[see][for a review]{2015ASSL..415..381G}, and are of importance for Space Weather.

Prominences are the limb counterpart of filaments, which when viewed on the disk, appear as dark material channelled along a polarity inversion line (PIL). The structure of filaments has been extensively studied and the role of the different attributes clearly identified (e.g. spine, feet, barbs). This description has led to a classification of filaments depending on the location (e.g. in active region, near the poles), on the shape, on the chirality and so one. Despite the extensive study for filaments, classification of prominences mostly relies on the projection of the structure on the plane-of-sky \citep[a historical review of morphological classification is presented in][]{1995ASSL..199.....T}. Recently, the small-scale structure of prominences has been investigated with an emphasize on the relationship between prominence material and cavities \citep{2011A&A...533L...1R, 2012ApJ...758L..37B}, and on the observed formation of cool material (e.g. condensation).

Classification of prominence can also be extended to their dynamics. Distinctions between active and eruptive prominences have been made in studies of the relationship of CMEs to prominences  \citep{2000ApJ...537..503G,2003ApJ...586..562G}. In these studies the radial or tangential motion of prominence dynamics is attributed to eruptive or active prominences, respectively. Directionality has also been employed by \citet{2015SoPh..290.1703M} in their extensive catalogue of filament eruptions, denoting radial, non-radial, and sideways (initial tangential component) eruptions. The author also documented other relevant attributes such as filament and eruption type (full/partial/confined), as well as structural indicators such as the twist, writhe, and symmetry. In a more recent study by \citet{2020JASTP.20505324Y}, the authors split eruptions into those from prominences and surges (a collimated outflow of plasma from a single source region), as well as distinguishing between different eruption speeds and latitudes. 

Methods to analyse prominence dynamics can be divided into those which employ line-of-sight (LOS) motions---using Doppler shifts from spectroscopic measurements---or those which are taken in the plane-of-sky (POS) (for a review see \cite{2015ASSL..415...79K}). LOS techniques provide direct measurements of plasma velocities along the viewing direction, but the constrained viewpoint can make it challenging to observe transverse motion. On the other hand, POS measurements are well-suited for tracking features (for example erupted prominence material), but are subject to projection effects and the temporal resolution of the data. 
However, full understanding of the detailed dynamics and structure may only be revealed through the aid of image processing techniques. We focus here on POS techniques which are able to discern structural or dynamical information, as this is the focus of our study. The OCCULT routine developed by \citet{2010SoPh..262..399A} is a robust method of detecting and tracing coronal loops in solar EUV and SXR observational data. Though OCCULT can detect a large amount of detected coronal features, the ridge-based detection can limits its ability to identify features that fall outside these criteria. Further techniques to enhance structure of solar images is found in. A more comprehensive list techniques of automated feature-detection algorithms for solar application can be found in \cite{2010SoPh..262..235A}.
A large portion of automated methods for discerning motion in the POS utilise optical flow techniques, such as local correction tracking \citep[LCT;][]{1988ApJ...333..427N}, or more recently, machine-learning techniques such as the DeepVel convolutional neural network \citep{2017A&A...604A..11A}. Comparisons of many of these techniques can be found in \citep{2007ApJ...670.1434W,2008ApJ...689..593C,2018SoPh..293...57T}. More recently \citet{2019A&A...624A..72Z} used optical flow techniques combined with doppler velocities to find the velocity of a CME-associated eruptive prominence.  

We introduce the Rolling Hough Transform (RHT) \citep{2014ApJ...789...82C, 2017SoPh..292..132S} to investigate the structural and dynamical behaviour of prominences. The RHT method has previously shown strong performance uncovering structural detail and orientation of EUV loops observed in on-disk active regions and chromospheric fibrils \citep{2017A&A...599A.133A,2017SoPh..292..132S}, as well in automated-timeslice measurements of coronal rain \citep{2017SoPh..292..132S,2022ApJ...931L..27S,2023ApJ...950..171S}. However, there are currently no such studies applying the technique to prominence dynamics. We present here a proof-of-concept for application of the spatial RHT algorithm to prominence dynamics by testing on two contrasting prominence events. Through this routine we are able to discern structural information and localised dynamics for each event and the different phases within. We construct a regime for classifying these events and phases as tangential or radial based on the structural orientation provided by the spatial RHT. Observational data for each event are presented in Section~\ref{sec:Sect2}. The methodology is explained in Section~\ref{sec:Sect3}, along with a detailed explanation of the spatial RHT algorithm. The results of the
spatial RHT for each event are provided in Section~\ref{sec:Sect4}, and the classification of prominence structure in Section~\ref{sec:Sect5}. Section~\ref{sec:Sect6} assesses the performance of spatial RHT algorithm. The results of spatial RHT applied to the prominence events are discussed in Section~\ref{sec:Sect7}, with concluding remarks about our study in Section~\ref{sec:Sect8}.

\section{Observations} \label{sec:Sect2}
\subsection{SDO/AIA Observations}
We study two distinct events to demonstrate the versatility of the Rolling Hough Transform algorithm. Observational data are taken from the 304 \AA\ channel from the Atmospheric Imaging Assembly \citep[AIA;][]{2012SoPh..275...17L} onboard the Solar Dynamics Observatory \citep[SDO;][]{2012SoPh..275....3P}. The 304 channel is dominated by the He \textsc{ii} emission at 304.8 \AA\ \citep{2010A&A...521A..21O} which corresponds to the cool prominence material \citep{2010SSRv..151..243L}. Prominence plasmas in this channel are optically thick, appearing bright against the dark background off-limb, allowing for clearer identification and observation of their structure. As with other EUV channels, there is a limb-brightening effect in the data  caused by the increasing line-of-sight path length toward the limb through the relatively optically thin corona. Whilst this can be corrected for and does not negatively impact our observations, it poses a problem later when we apply the spatial RHT algorithm (see details in Section~\ref{sec:Sect4}). AIA has an effective spatial resolution of 1\farcs5 (0\farcs6 per pixel), allowing for detailed observations of prominences. The fine temporal cadence of 12 seconds enables the instrument to capture fine-scale motion involved in prominence dynamics in near real-time. 
 
\subsection{Quiet-Sun Eruption}\label{sec:Sect2.2}
\begin{figure*}[t]
\graphicspath{ {./images/} }
    \centering
    \includegraphics[width=\linewidth]{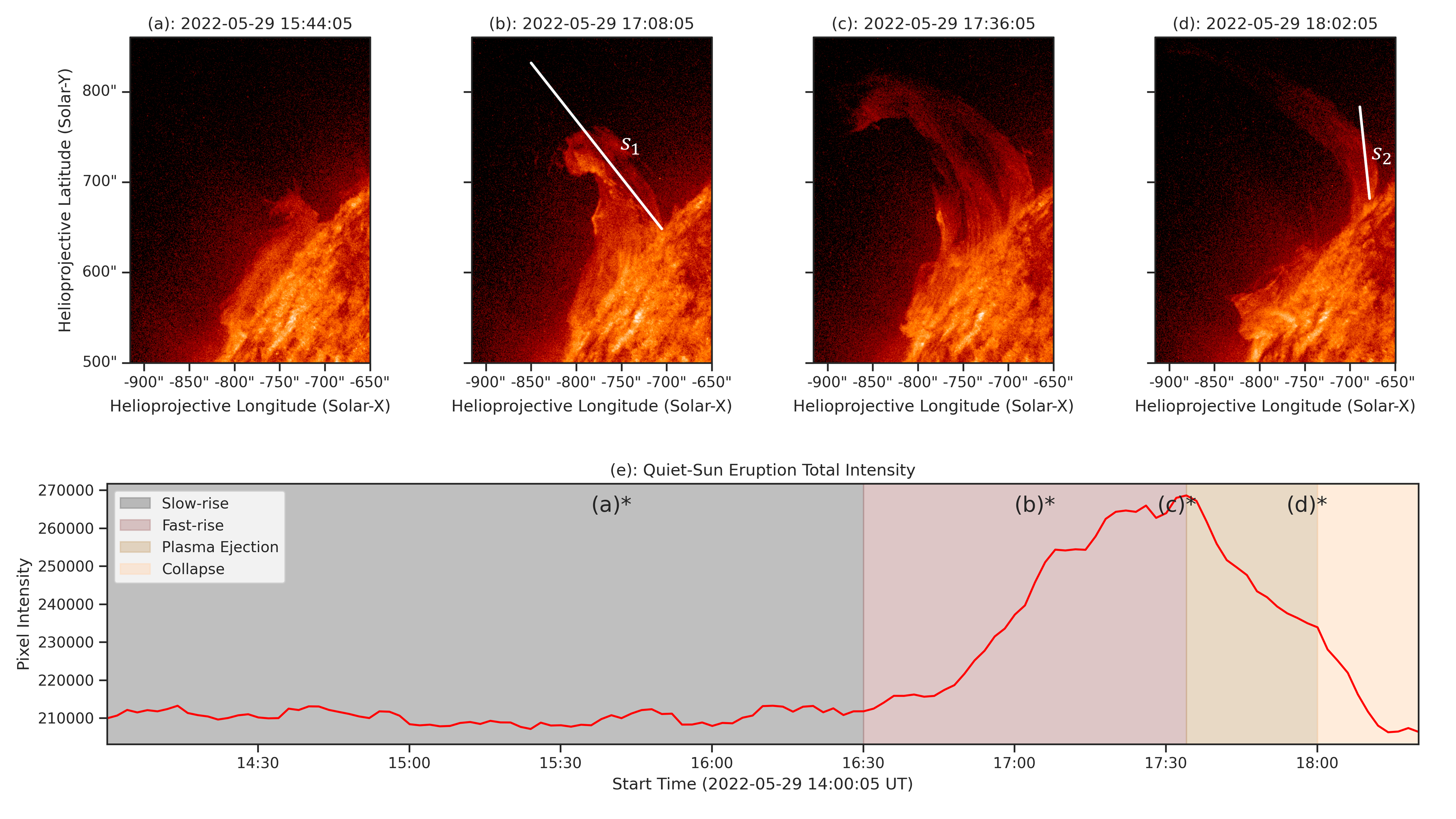} 
    \caption{Top: SDO/AIA 304 \AA\ images of the quiet-Sun eruption across each of the four phases: (a) Pre-eruption, (b) Eruption, (c) Plasma Ejection, and (d) Collapse. Bottom: (e) Light curve for the quiet-Sun eruption generated across the  full timeseries, coloured according to each phase. The positions in time of the above SDO/AIA images are marked with an asterisk. An animation (Movie1) of the SDO/AIA 304 \AA\ observations for the quiet-Sun eruption is available online. The animation is taken from 14:00 UT to 18:20 UT, with a cadence of 2 minutes and runs at 30 frames-per-second. Details of the phases of the event are found within Section~\ref{sec:Sect2.2}.}
    \label{fig:Fig1}
\end{figure*}

The quiet-Sun eruption was observed on 2022 May 29 between 14:00 UT and 18:20 UT on the East solar limb. We obtained a timeseries of this event with a 2 minute cadence. In Figure~\ref{fig:Fig1}, we provide 4 snapshots (top panel) that are characteristic of the different evolution stages identified in the light curve of the total intensity in the field-of-view of interest (bottom panel). An animation of this figure is also available online (Movie1). The start of the prominence activation is observed at 14:00 UT (Figure~\ref{fig:Fig1}(a)), with the slow rise of the prominence. The prominence begins rapidly accelerating around 16:30 UT, forming the fast-rise phase (Figure~\ref{fig:Fig1}(b)). This two-phase evolution of an initial slow rise followed by a sudden transition to a fast rise has been described in numerous other studies \citep[e. g.][]{2005ApJ...630.1148S,2006A&A...458..965C,2013ApJ...778..142T}. \citet{2008ApJ...674..586S} included the sudden transition, whereby there is a rapid acceleration of the prominence, as a further eruption phase. For the event studied here, we find it useful to define two further phases, in addition to the slow and fast-rise phases, for the remainder of the eruption: the ejection phase, beginning at 17:45 UT and coinciding with the ejection of plasma material (Figure~\ref{fig:Fig1}(c)), and the collapse phase at 18:00 UT, whereby the ejected material plasma drains back to the solar surface (Figure~\ref{fig:Fig1}(d)). 

The 4 phases of the eruption event are denoted by the different coloured regions in the light curve of Figure~\ref{fig:Fig1}(e), and are defined by sudden changes in the intensity gradient. We only consider intensity from pixels off the limb by masking out the solar disk. During the pre-eruptive phase there is little change in the intensity, forming a baseline. At 16:30 UT we observe a transition period towards a fast rise (beginning at 16:40 UT and coinciding with a large increase in intensity). A similar transition has been reported by \citet{2006A&A...458..965C}. Brightenings are observed at this time around the top of the prominence body. The peak of intensity corresponds with the onset of plasma ejection, after which it decreases. A further sharp decrease in intensity results from the remaining material draining back to the chromosphere.

In general, the nature of the prominence eruption --- whether partial, failed (confined), or otherwise --- remains uncertain \citep{2007SoPh..245..287G}, as the fraction of prominence material that successfully escaped is unknown. Although for this example, observational evidences suggest that it was a failed eruption characterised by the importance of the drainage phase, and the lack of evidence of the top of the prominence leaving the AIA field-of-view. Using the CACTUS database for CMEs \citep{2009ApJ...691.1222R}, we identify two possible CME candidates with a similar time and location on the solar limb, though their connection to the events observed cannot be confirmed.  

\subsection{Polar-swirl Activation}\label{sec:Sect2.3}
\begin{figure*}[t]
\graphicspath{ {./images/} }
    \centering
    \includegraphics[width=\linewidth]{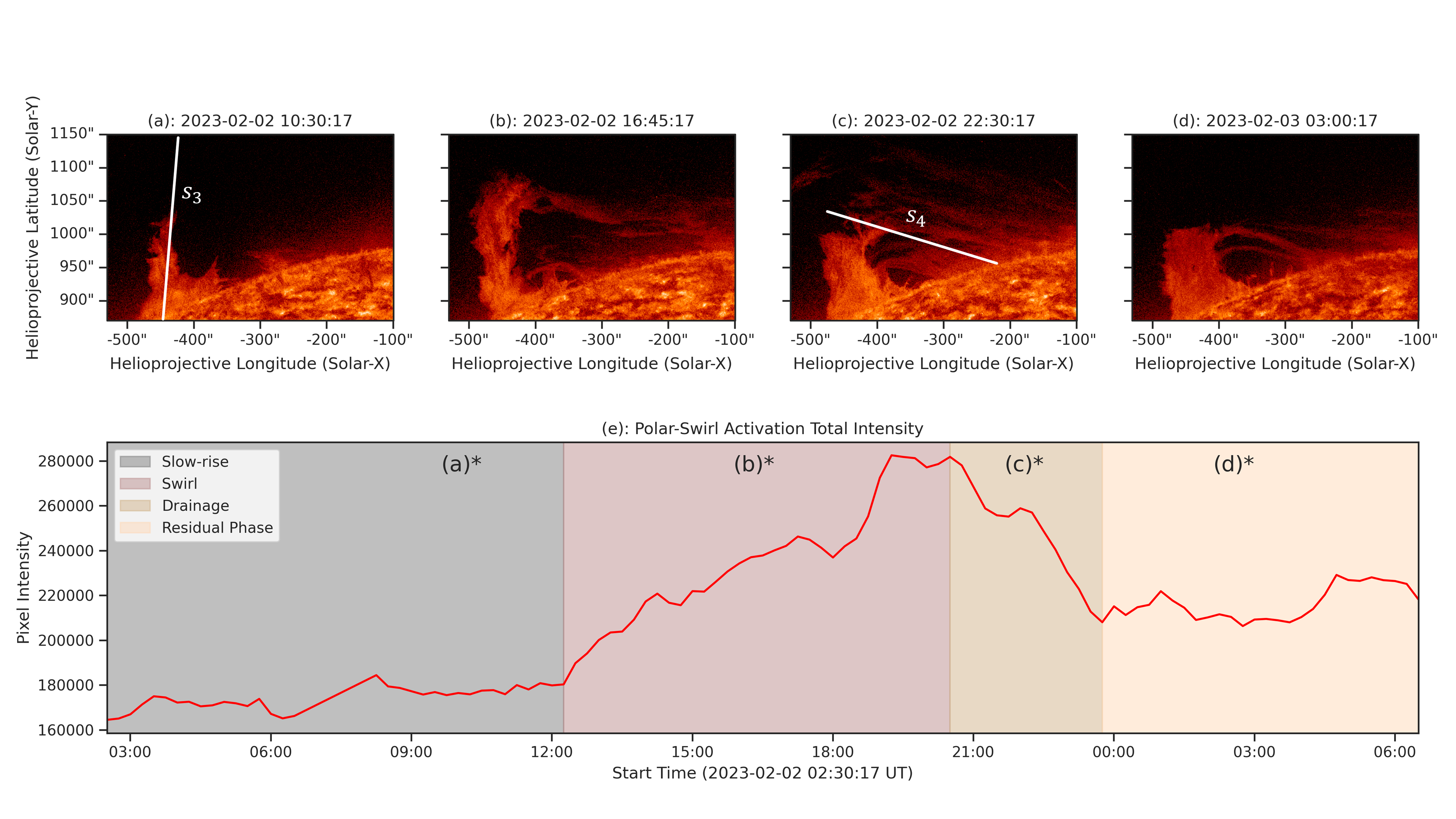} 
    \caption{Top: SDO/AIA 304 \AA\ images of the polar-swirl activation across each of the four phases: (a) Rise, (b) Swirl, (c) Drainage, and (d) Residual Phase. Bottom: (e) Light curve for the polar-swirl activation generated across the  full timeseries, coloured according to each phase. The positions in time of the above SDO/AIA images are marked with an asterisk. An animation (Movie2) of the SDO/AIA 304 \AA\ observations for the polar-swirl activation is available online. The animation is taken from 2023-02-02 02:30 UT to 2023-02-03 06:30 UT, with a cadence of 15 minutes and runs at 15 frames-per-second. Details of the phases of the event are found within Section~\ref{sec:Sect2.3}.}
    \label{fig:Fig2}
\end{figure*}

A polar crown prominence was observed close to the solar North pole (above a latitude of about 60 degrees) on 2023 February 02 02:30 UT and lasted to to 2023 Feb 03 06:30 UT\footnote{The full event is observed lasting for several days, however, we only consider the period of greatest activity for our study here}. We use a cadence of 15 minutes and focus only on the East-side of the event, where the majority of the prominence dynamics are observed. In Figure~\ref{fig:Fig2}, we present 4 snapshots (top) that are characteristic of the different evolution stages in the light curve of the total intensity in the field-of-view of interest (bottom). An animation of this figure is also available online (Movie2). The start of prominence activation is observed at 02:30 UT (Figure~\ref{fig:Fig2}(a)), with the slow rise of prominence on the East limb (latitude of 64 degrees), appearing as a column in the plane-of-sky. The plasma reaches an initial maximum height at around 11:30 UT, and then begins to curve almost 90 degrees laterally to the Sun's northern pole as the system destabilises (Figure~\ref{fig:Fig2}(b)). We observe flows moving both towards and away from the prominence, originating from various high-latitude points. The transferred plasma is swirling around the northern pole, hence we term this the swirl phase. At 20:30 UT (Figure~\ref{fig:Fig2}(c)), the upper half of the prominence column breaks away and is assimilated with other activated plasma in what we term the drainage phase. From 23:45 UT onwards (Figure~\ref{fig:Fig2}(d)) there is a decrease in activity, and the observed prominence body slowly recedes in height. This residual mass is still observed several days after the eruption. Plasma motion is still observed in this residual phase, but appears to be travelling behind the limb. 

The 4 phases of the polar swirl activation are captured in the light curve of Figure~\ref{fig:Fig2}(e), and are defined by sudden changes in the intensity gradient. The solar disk is masked out as to only consider off-limb pixels. The event begins with the slow-rise phase where the intensity is largely stable. In contrast to the quiet-Sun eruption, there is no fast-rise phase or sudden acceleration of the prominence. We observe an increase in the intensity at 12:15 UT, but it is instead correlated with the destabilisation of the prominence and the swirling of plasma material. The drainage phase is initiated at 20:30 UT when the upper half of the prominence breaks away and, along with other plasma, is drained down along field lines to the chromosphere. Consequently, the intensity in this phase decreases as plasma leaves the field-of-view. 23:45 UT marks the residual phase and sees a return to relative stability in the intensity and dynamics, at least in the line-of-sight.

The polar swirl event represents a more unique eruption case, and to our knowledge, no other studies of this particular event have been carried out. Due to the large scale of this event, and with only one point-of-view, it is difficult to assess the detailed mechanisms behind the dynamics, although we can surmise that this is another example of a failed eruption. The rising prominence on the East limb characteristically begins with an initial slow rise which is interrupted through interactions closer to the pole. The destabilisation of the prominence seems to prevent the eruption process and instead causing the swirling motion to dominate the dynamics of the event. Such an event is perhaps better described as an prominence activation. \citet{2000ApJ...537..503G} describes active prominences as prominences which exhibit motion but do not appear to escape the solar gravitational field. The predominant motion is tangential, or lateral to the surface: the prominence material is transported parallel to the limb, rather than ejected. 

\subsection{Kinematics}
The analysis of the kinematics for both events was carried out using timeslice techniques. In particular, we examine the motion during the different phases of the events as defined in Sections~\ref{sec:Sect2.2} and~\ref{sec:Sect2.3}. We use structural detail observed in the movies to determine the most optimal position of the slits. All the speeds reported below are measured in the POS.

\begin{figure}[h!]
\graphicspath{ {./images/} }
    \centering
    \includegraphics[width=\linewidth, trim=50 0 50 0, clip]{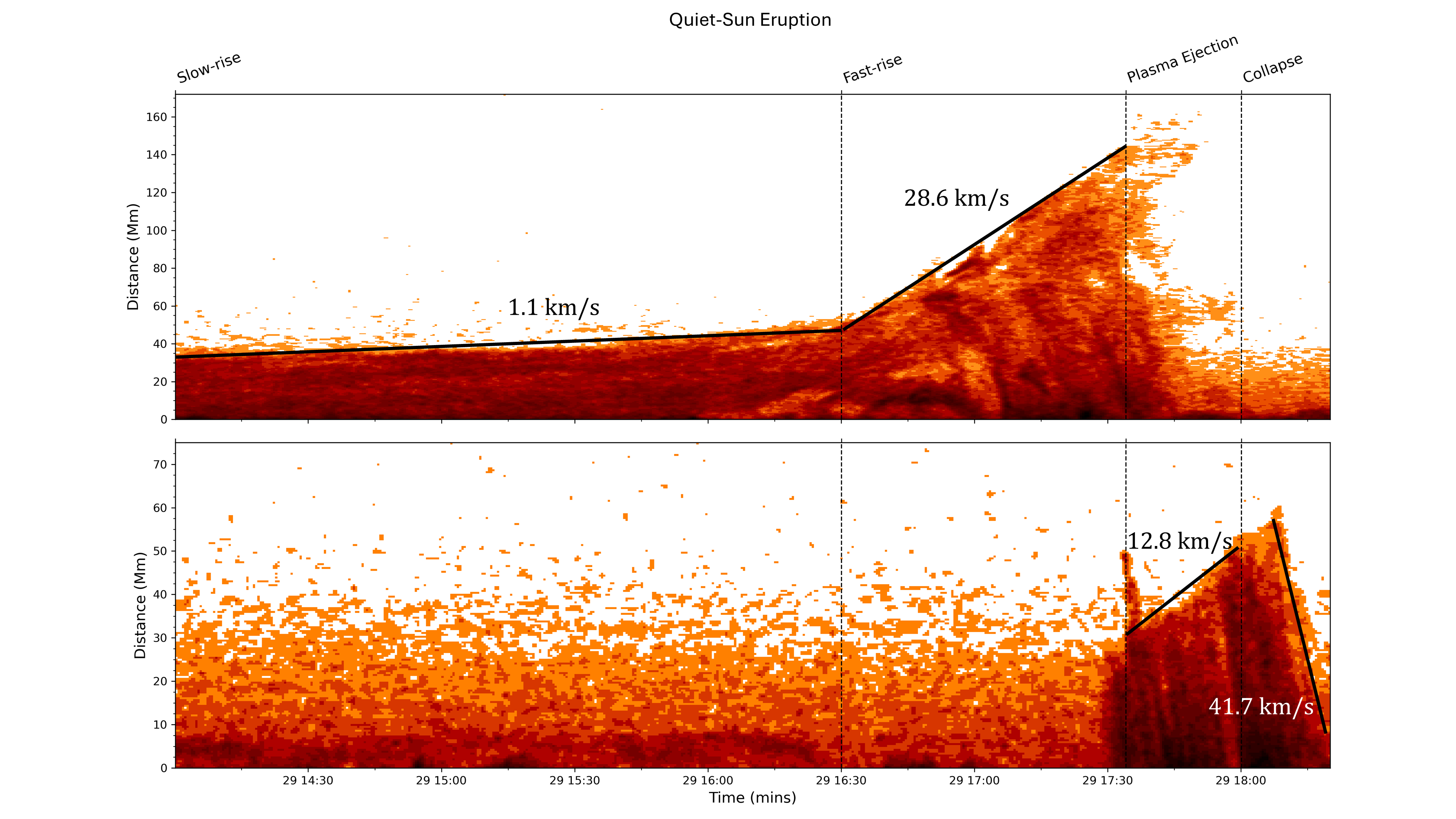} 
    \caption{Timeslice for the quiet-Sun eruption. Top: timeslice taken along slice $s_1$, used to extract the speed of the slow-rise and fast-rise phases. Bottom: timeslice taken along slice $s_2$, used to extract the drainage speed.}
    \label{fig:Fig3}
\end{figure}

Figure~\ref{fig:Fig3} shows the timeslices for the quiet-Sun eruption, with the top panel showing slice $s_1$ (Figure~\ref{fig:Fig1}(b)) for the slow-rise and fast-rise speeds, and the bottom panel slice $s_2$ (Figure~\ref{fig:Fig1}(d)) for the plasma ejection and drainage speeds. The speeds are collated in Table~\ref{tab:Tab1}. Our findings show strong agreement with previous studies regarding the slow- and fast-rise speeds of prominence eruptions \citep{2011A&A...533L...1R,2015SoPh..290.1703M}, as well as for the drainage drainage phase \citep{2012ApJ...745L..21L}.

\begin{table}[htbp!]
\centering
\caption{Characteristic speeds for the quiet-Sun eruption}
\label{tab:Tab1}
\begin{tabular}{c c}
\hline
\hline
Event & Speed (\kms) \\
\hline
Slow-Rise & 1.1 \\
Fast-Rise & 28.6 \\
Plasma Ejection & 12.8 \\
Drainage  & 41.7 \\
\hline
\end{tabular}
\end{table}

\begin{figure}[h!]
\graphicspath{ {./images/} }
    \centering
    \includegraphics[width=\linewidth, trim=50 0 50 0, clip]{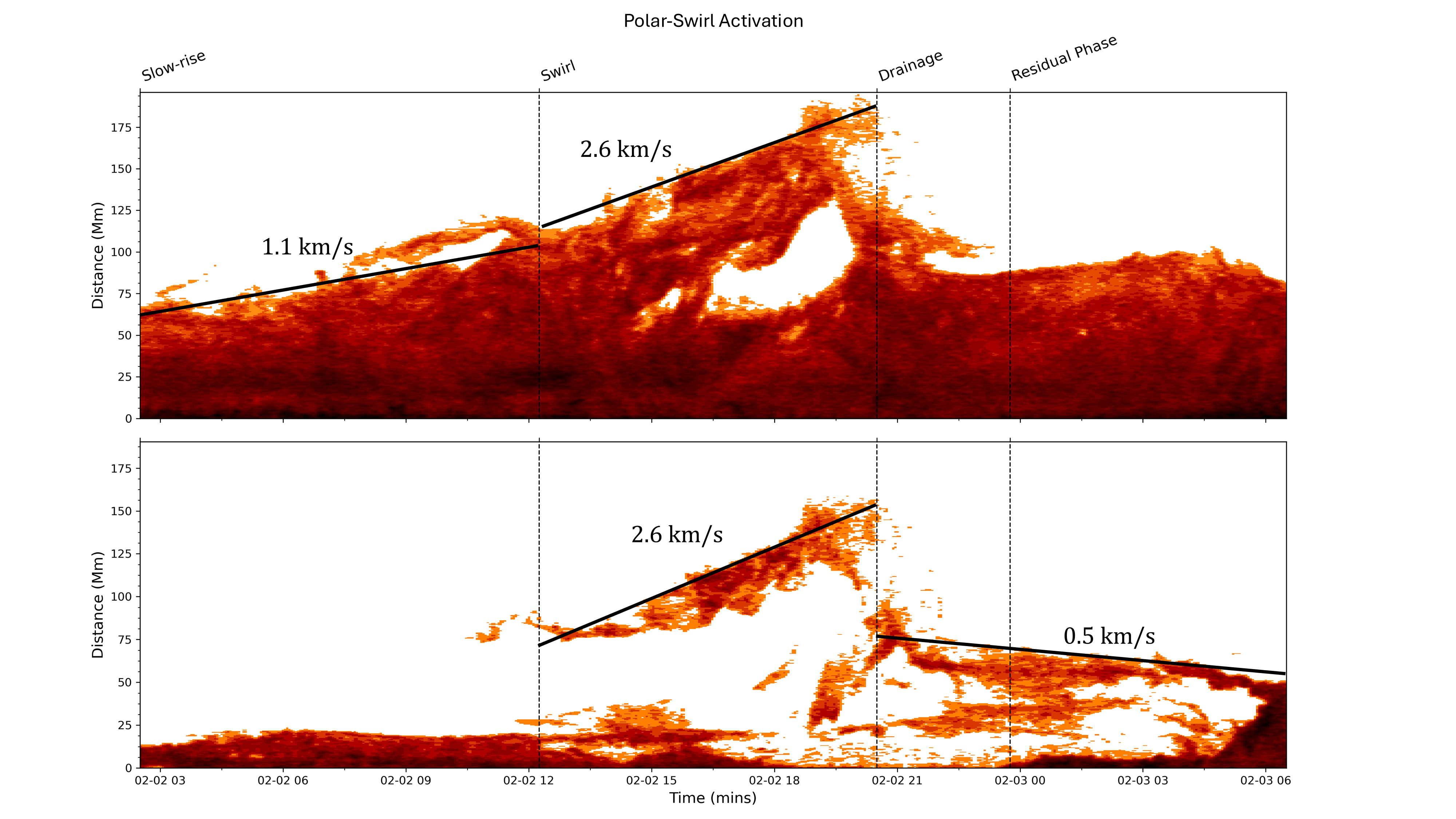} 
    \caption{Timeslice for the polar-swirl activation. Top: timeslice taken along slice $s_3$, used to extract the speed of the slow-rise phase and part of the swirl. Bottom: timeslice taken along slice $s_4$, used to extract the swirl and drainage speeds.}
    \label{fig:Fig4}
\end{figure}

Figure~\ref{fig:Fig4} shows the timeslice for the polar-swirl activation. The top panel shows slice $s_3$ (Figure~\ref{fig:Fig2}(a)) from which we obtain estimates of the slow-rise and swirl speed. The bottom panel shows slice $s_4$ (Figure~\ref{fig:Fig2}(c)), from which we extract a second estimate of the swirl speed along with part of the drainage velocity. These speeds are collated in Table~\ref{tab:Tab2}. The kinematics in this event are more complicated due to the more complex dynamics, especially during the swirl phase where there is a large amount of diffuse plasma obscuring activity close to the limb. Nevertheless, these results show striking similarity in the initial slow-rise phase of both events, as they share the same speed. However, the polar swirl event diverges relatively early, with its activation speed being only slightly higher, suggesting significantly lower acceleration compared to the fast-rise phase of the quiet-Sun eruption. Additionally, the speeds observed during the swirl and drainage phases are comparable to those throughout the rest of the event, indicating generally slower motion for the polar swirl event overall. 

\begin{table}[htbp!]
\centering
\caption{Characteristic speeds for the polar-swirl activation}
\label{tab:Tab2}
\begin{tabular}{c c}
\hline
\hline
Event & Speed (\kms) \\
\hline
Slow-Rise & 1.1 \\
Swirl     & 2.6 \\
Drainage  & 0.5 \\
\hline
\end{tabular}
\end{table}

\section{Spatial RHT}\label{sec:Sect3}
\begin{figure*}[t]
\graphicspath{ {./images/} }
    \centering
    \includegraphics[width=\linewidth, trim=10 200 10 10, clip]{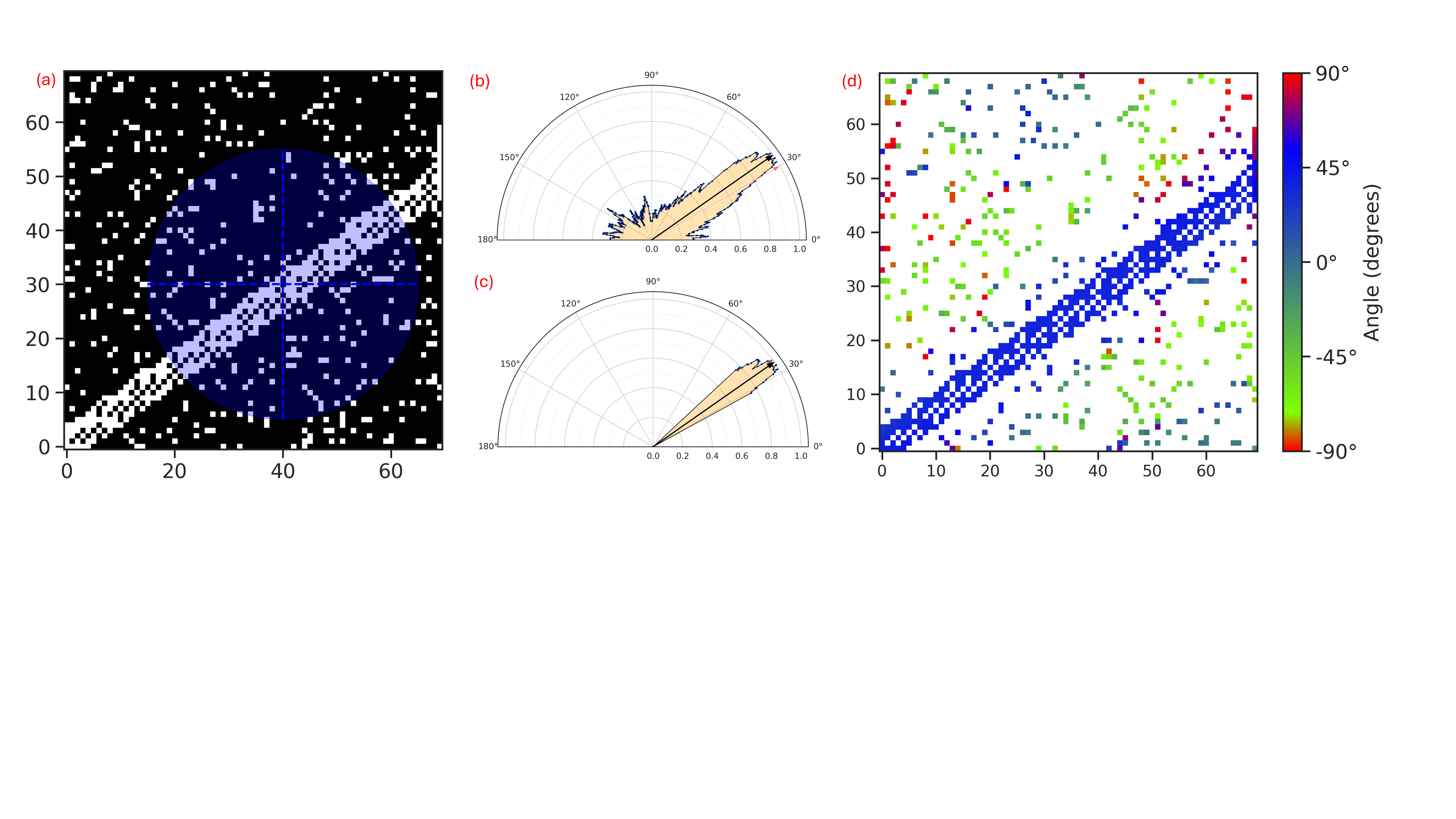} 
    \caption{(a) Bitmap array with a line at 35 degrees from the horizontal axis and random noise. The kernel is shown centered on pixel $(x,y)$ covered by the blue circular region. For each angle bin, the number of pixels which are linearly aligned is counted and normalised. This is shown in the polar plot (b), with the black angle showing the true line angle and the red arrow the mean angle. (c) Shows the effect of thresholding the bin values to only the top 75\% resulting in a better agreement with the true angle. (d) Final spatial RHT output of the bitmap for all pixels.}
    \label{fig:Fig5}
\end{figure*}

The Rolling Hough Transform (RHT) is an algorithm for identifying and classifying the spatial orientation of curvilinear features within an image using directional statistics \citep{mardia1999directional}. We use here the version of the code adapted for solar physics by \citet{2017SoPh..292..132S}, although other versions of the algorithm exist.
The goal of the algorithm is to find the function $H_{xy}(\theta)$ for each pixel $(x,y)$ in a timeseries of images. The full procedure is as follows: 

\begin{enumerate}[label=(\arabic*)]
    \item The input array is spatially filtered to enhance the features of interest. A bitmap is thus generated with features of interest taking a value of 1, as seen in Figure~\ref{fig:Fig5}(a).
    \item A circular window (kernel) of diameter $D_{W}$ is created around each image-space pixel $(x,y)$ (blue circle in Figure~\ref{fig:Fig5}(a)).
    \item The Hough transform is performed for $\rho = 0$ (only lines passing through pixel origin $(x,y)$), giving $H_{xy}(\theta)$. In other words, it calculates the number of linear, aligned pixels around each angle bin within the circular kernel that pass through the centre pixel. The number of angle bins is defined by the kernel width.
    \item An adaptive threshold, $f$, is defined as a percentage of max[$H_{xy}(\theta)$]: only $\theta$ values above $f$ will be stored, the rest being set to zero. The thresholding alleviates the influence of random noise on the measured angle (see comparison of Figures \ref{fig:Fig5}(b) and (c)). The value of $f$ is set to 0.25 \citep[see][]{2017SoPh..292..132S}: the choice of this value produces good results for the timeseries used in the next sections.
    \item Three outputs are computed. We determine the mean axial direction, $\theta_{xy}$ for pixel $(x,y)$, henceforth referred to as the mean angle, using the directional statistics of \citet{mardia1999directional},
    \begin{equation}
        \bar{\theta}_{xy} = 0.5  \arctan(y/x)~\textrm{.}
    \end{equation}
    We convert the angle output from 0 to 2$\pi$ radians into a range from -90 to 90 degrees. This represents angles within the image ($xy$) plane, with angles $\pm 90$ degrees corresponding to vertical features (parallel/anti-parallel with the $y$-axis), and 0 degrees corresponding to horizontal features (parallel with the $x$-axis). The error in the mean angle is given by $\epsilon_{0}$, calculated from Equation (8) in \citet{2017SoPh..292..132S}. And the third output is the mean resultant length, $\bar{R}$, defined as 
    \begin{equation}
        \bar{R} = \sqrt{\bar{C}^{2}+\bar{S}^{2}}~\textrm{,}
    \end{equation}
    where $\bar{C}$ and $\bar{S}$ are the weighted components of the RHT distribution (see equations (3) and (4) of \citet{2017SoPh..292..132S}). $\bar{R}$ takes values from 0 to 1 and represents the pixel (feature) connectivity/dispersion. Thus filtering for higher $\bar{R}$ leads to more coherent structure in the resulting image. 
    \item The steps (2)--(5) are repeated across all non-zero pixels in the bitmap, storing each of the three outputs in separate arrays (see Figure~\ref{fig:Fig5}(d) for the spatial RHT mean angle output). The mean angle may be thresholded using max[$H_{xy}(\theta)$], $\bar{R}$, and $\epsilon_{0}$. Following the analysis by \citet{2017SoPh..292..132S}, we threshold the spatial mean angles to include only values 
    satisfying $\bar{R} \geq 0.75$, which alleviates cases where there may be crossing features. For a timeseries, these steps are applied across each spatial slice in the array sequentially.

\end{enumerate}

The spatial RHT routines includes further parameters to control the noise level, \texttt{nlev}, and the strength of the median filter, \texttt{smr\_xy}. These can can be fine-tuned to improve the final spatial RHT output. A full description of these parameters, along with our values and notes on setting appropriate values, is found in Table \ref{tab:Tab5} of Section~\ref{sec:Sect6}. Other factors also of importance to the spatial RHT algorithm, notably the quality of the input image and the features within, are also discussed. For a timeseries we compute a global mean angle by averaging over time. The impact of using a timeseries input for the spatial RHT and subsequent analysis is discussed in Section~\ref{sec:Sect6}.

 We validate the spatial RHT results by reproducing the TRACE image 171 \AA\ image of NOAA AR 08222 from \citet{2017SoPh..292..132S} (Figure 4(d)). The results confirmed that the code was working as intended, with both the detected RHT features and angles matching with Schad's result.

\section{Prominence structure and evolution from spatial RHT} \label{sec:Sect4}
\subsection{SDO/AIA Frame of Reference}
To facilitate more detailed analysis of solar image data, we convert the spatial RHT angles from the image plane ($xy$) into polar angles (radial and tangential). For each pixel ($x$,$y$) in the field-of-view of the event, we define reference angle $\alpha$, representing the angle between the line connecting pixel ($x$,$y$) to the origin (0,0)(the centre of the solar disk), and the horizontal (east-west) direction. The angle $\alpha$ is measured anticlockwise from the positive x-axis (eastward direction) and ranges from 0 to $2\pi$. We then apply the following modification on the RHT angle, $\bar{\theta}_{xy}$, to obtain the polar angle, $\bar{\theta}_{polar}$,
\begin{equation}
    \bar{\theta}_{polar} = \alpha - \bar{\theta}_{xy}~\textrm{.}
\end{equation}
Additionally, we restrict the result to a range of $0$ to $\pi$ through the modulo operation ($mod(\pi)$) and offset by $-\pi$ convert to the desired range of $-90$ to $90$ degrees. In this frame, radial structures are $\pm90$ degrees and are perpendicular to the solar limb (shown in dark purple in our results). Tangential structures are $0$ degrees and are parallel to the limb (shown in grey in our results). Angles in-between these limits consist of both radial and tangential components, and can be consider strongly/weakly radial/tangential structures. The definition of positive or negative radial angles, can be understood as representing features with anti-clockwise or clockwise directionality, respectively, relative to their originating vector at the solar limb. We highlight this directionality in our RHT results by colouring clockwise features in blue, and anti-clockwise features in red.

The RHT produces a contribution from the solar limb, as this is interpreted as an edge in the image. Limb brightening from EUV channels does exacerbate this effect, by enhancing the width of the detected edge, but removing the brightening will not entirely resolve the issue, as a line will inevitably be detected at any sharp intensity gradient or boundary within the image. The detected limb edge needs to be removed to prevent tangential angles from dominating distributions. By modifying the masking function used remove the disk for the light curves, we can remove these unwanted pixels, setting a threshold which varies the solar radius. This procedure does not result in a significant loss of information, as most of the structural detail is located above the limb rather than along its immediate boundary. 

We present the spatial RHT results for each event, first showing the global properties across the full timeseries, and then the results from each phase to quantify more local dynamics. All results shown are averages across each individual timeseries, rather than individual slices. The average error obtained by the RHT error for these results was around 5 degrees.

\subsection{Quiet-Sun Eruption}
\begin{figure}[h]
\graphicspath{ {./images/} }
    \centering
    \includegraphics[width=\linewidth]{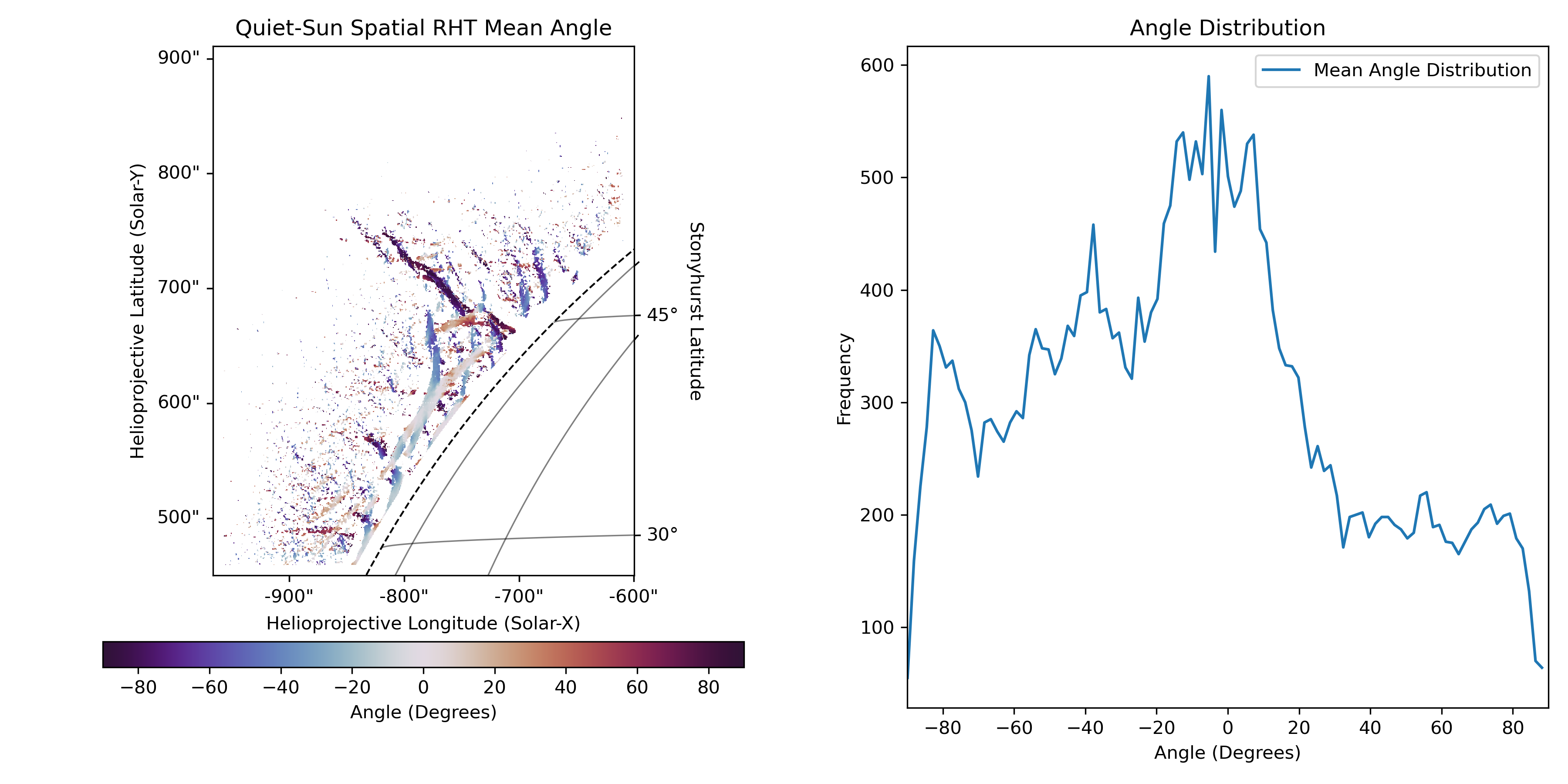} 
    \caption{Spatial RHT mean angle for the quiet-Sun eruption (left) with the distribution of angles (right). Angles of $\pm90$ and 0 degrees represent radial and tangential features, respectively.}
    \label{fig:Fig6}
\end{figure}

Figure~\ref{fig:Fig6} shows the spatial RHT mean angle result applied to the full timeseries for the quiet-Sun eruption (left), along with the extracted distribution (right). The cadence used for these results was 1 minute. The spatially RHT shows the tangentially oriented plasma body (in grey), present during the slow-rise phase (Figure~\ref{fig:Fig1}(a)). Above this are features corresponding to the fast-rise and ejection phases, namely the eastern-edge of the prominence (in blue) and the radial line (in purple) of the ejected plasma (Figure~\ref{fig:Fig1}(b) and (c)). The RHT also uncovers the lower sections of drainage pathways throughout the image, oriented radially clockwise (in blue). This information is reflected in the distribution, with the dominant peak around 0 degrees corresponding to the primary tangential structure. A second peak around -30 degrees is associated with the drainage pathways, while a third peak at -90 degrees represents the radially expanding plasma.

\begin{figure*}[t]
\graphicspath{ {./images/} }
    \centering
    \includegraphics[width=\linewidth]{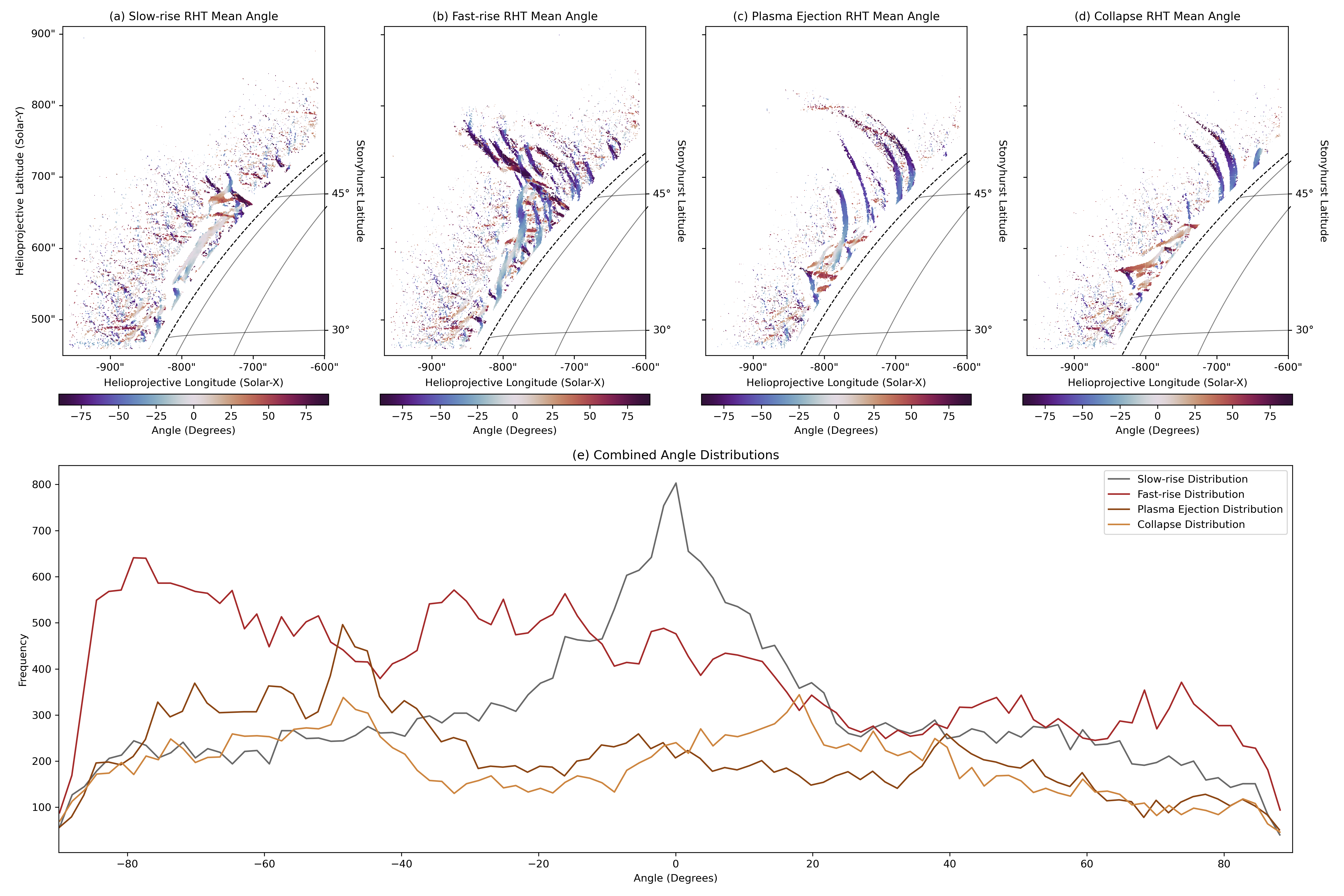} 
    \caption{Top: Spatial RHT mean angles for the four phases of the quiet-Sun eruption: slow-rise (a), fast-rise (b), plasma ejection (c), and collapse (d). Bottom: (e) Angle distributions for each phase in the quiet-Sun eruption. Angles of $\pm90$ and 0 degrees represent radial and tangential features, respectively.}
    \label{fig:Fig7}
\end{figure*}

In Figure~\ref{fig:Fig7}, we show the spatial RHT mean angle results for each of the 4 phases (top) along with the corresponding angle distributions (bottom). We use a cadence of 1 minute for the timeseries data for each phase. Compared with the full timeseries we now see more distinct and detailed features from each phase. The spatial RHT of the slow-rise phase (Figure~\ref{fig:Fig7}(a)) shows the the accumulation of the tangentially oriented material on the limb, which grows into a radial column plasma during the fast-rise phase (Figure~\ref{fig:Fig7}(b)). This column shows many fine structural features, which comprise of both material expanding out of the corona and material falling back towards the solar surface. Several of the drainage paths, whereby the majority of the mass is transferred back to the chromosphere during the plasma ejection and collapse phases, have been fully isolated by the RHT and are largely radially oriented (Figure~\ref{fig:Fig7}(c) and (d)). Some of this mass drainage begins to re-accumulate on a lower latitude point of the structure, detected by the RHT as a anticlockwise-radial structure (in red).

The structural information for each phase is corroborated in the angle distribution in Figure~\ref{fig:Fig7}(e). The slow-rise phase shows a dominant peak around $0$ degrees, corresponding to the accumulated plasma on the limb. This material evolves radially outwards, as seen by the peaks at $\pm 90$ degrees in the fast-rise distribution. The remaining two phases show less distinctive peaks but nonetheless display a range of quasi-radial alignment, a product of the curvilinear drainage pathways and the re-accumulated mass lower down the structure.

\subsection{Polar-swirl Activation}
\begin{figure}[t]
\graphicspath{ {./images/} }
    \centering
    \includegraphics[width=\linewidth]{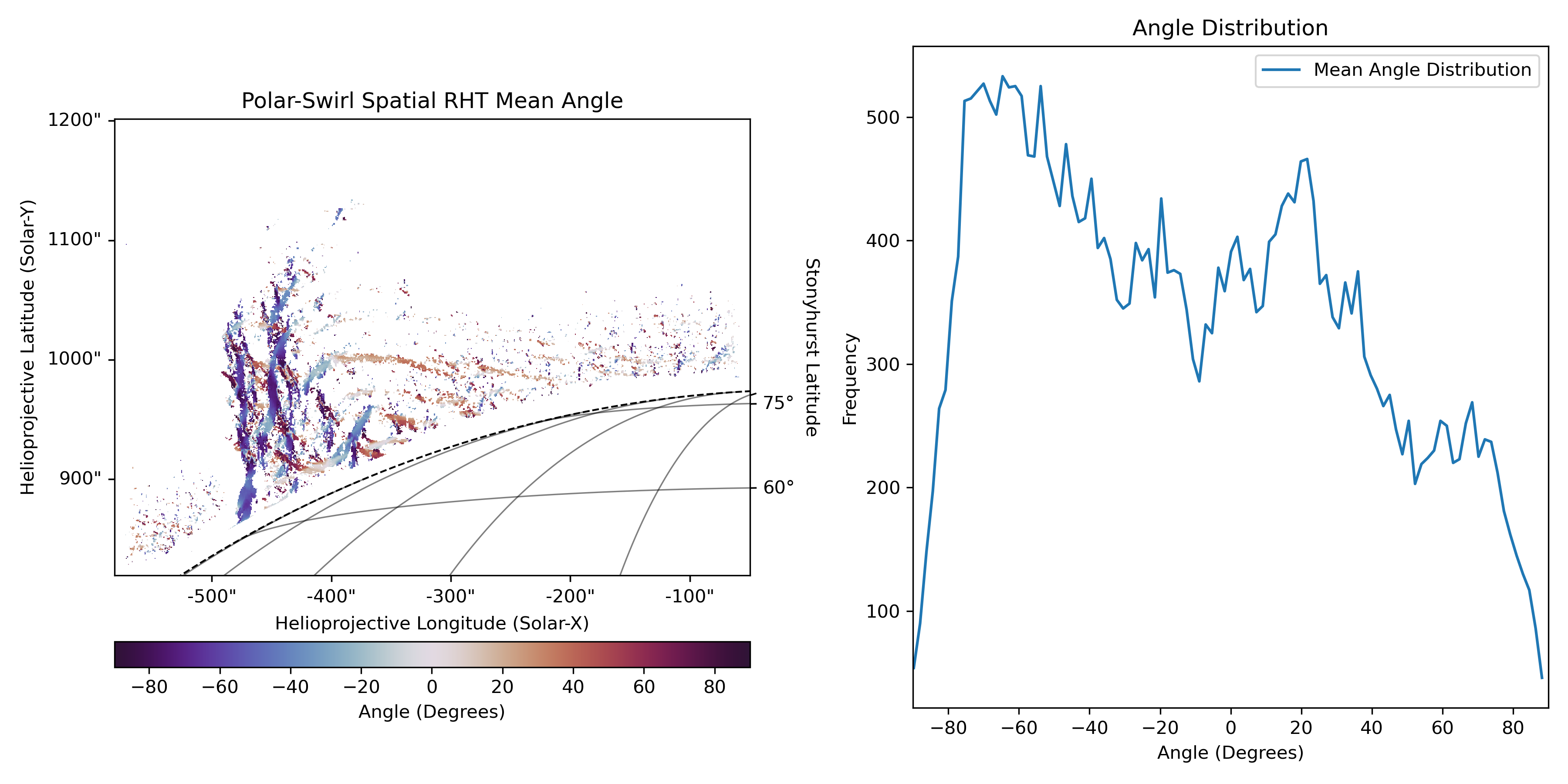} 
    \caption{Spatial RHT mean angle for the polar-swirl activation (left) with the distribution of angles (right). Angles of $\pm90$ and 0 degrees represent radial and tangential features, respectively.}
    \label{fig:Fig8}
\end{figure}

Figure~\ref{fig:Fig8} shows the spatial RHT mean angle result applied to the full timeseries for the polar-swirl (left) and the corresponding angle distribution (right), using a cadence of 5 minutes. Much of what the RHT has detected is from the prominence body, as much of this material is present throughout the whole event. Within the body are features oriented clockwise (in blue), reflecting the transverse `swirling' of the plasma during activation (Figure~\ref{fig:Fig2}(b)). Anti-clockwise features (in red) can be spotted at higher latitudes in the drainage phase (Figure~\ref{fig:Fig2}(c)) and are continuations of the arching plasma paths as they curve back down to the chromosphere, reversing orientation. These paths are sparsely connected as they lie within a cloud of diffuse material which is swept up above the limb during the event. The distribution shows a wide range of angles with a strong clockwise (negative) radial angle bias, resulting from features comprising the prominence body, with a steady decline as angles move towards $+90$ degrees. A second peak near the tangential at 20 degrees represents the change in direction from the curving plasma lines from the drainage phase.

\begin{figure*}[t]
\graphicspath{ {./images/} }
    \centering
    \includegraphics[width=\linewidth]{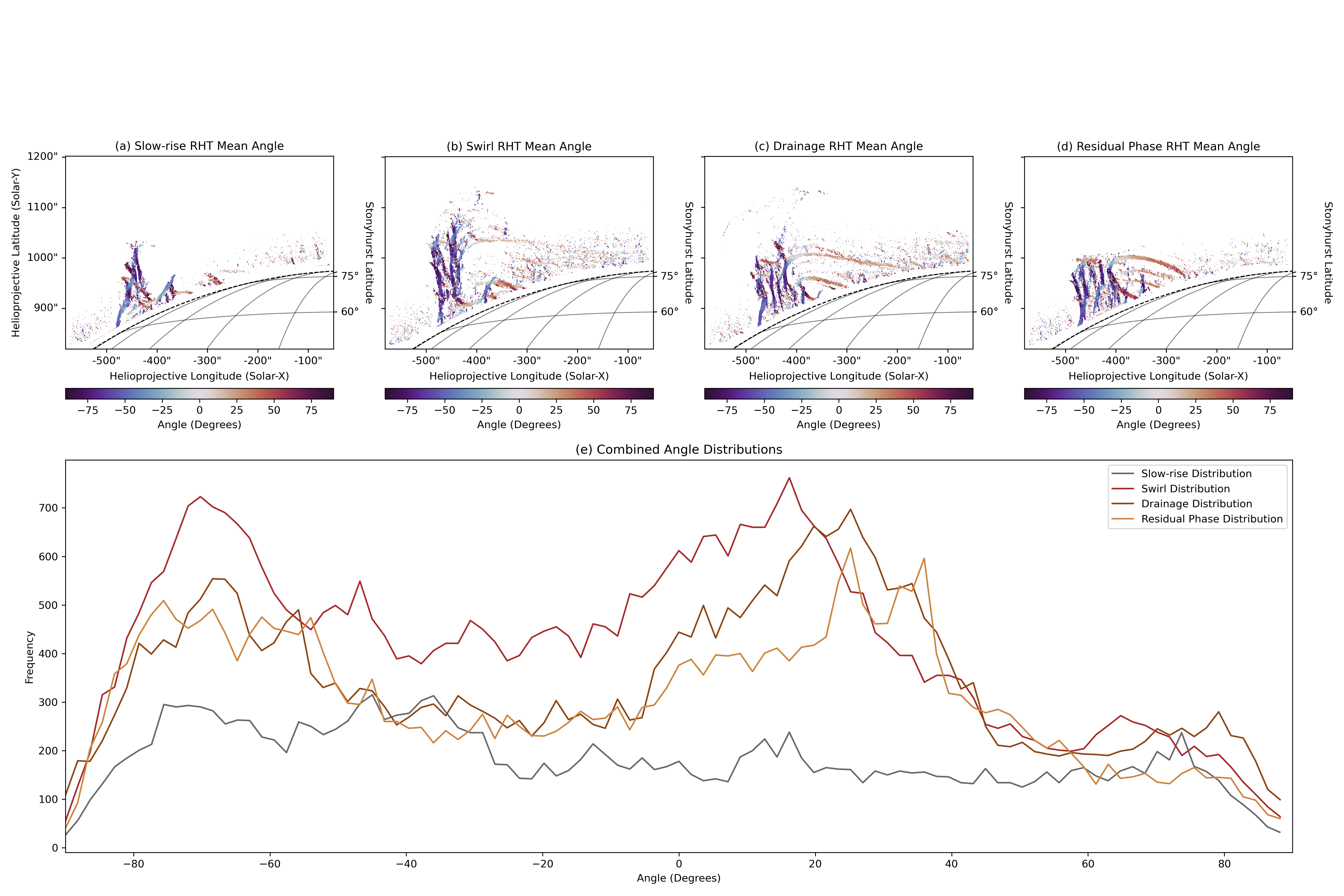} 
    \caption{Top: Spatial RHT mean angles for each of the four phases: fast-rise (a), swirl (b), drainage (c), and the residual phase (d). (e) Angle distributions for each phase in the polar-swirl activation. Angles of $\pm90$ and 0 degrees represent radial and tangential features, respectively.}
    \label{fig:Fig9}
\end{figure*}

In Figure~\ref{fig:Fig9}, we show the spatial RHT mean angle results for each of the 4 phases (top) along with the corresponding angle distributions (bottom). We use a cadence of 10 minutes for the timeseries data for each phase. Similar to the quiet-Sun results, the spatial RHT shows an accumulation of material on the east-limb during the slow-rise phase (Figure~\ref{fig:Fig9}(a)), although the structure here is more radially oriented than tangential. As the prominence rises radially, we begin to see the interactions of the plasma with features closer to the solar north pole, appearing as weakly-radial features on the right hand side of the prominence (Figure~\ref{fig:Fig9}(b)). The RHT also uncovers a strong drainage line near the prominence footpoint. Figure~\ref{fig:Fig9}(c) reveals the detailed structure of the lower prominence body, with numerous features detected by the RHT across a range of orientations. Radial features within the lower prominence body indicate that  the general structure has been retained, but is permeated by drainage pathways which arch towards higher latitude points, displaying a greater range of angles. The top of the prominence structure is disrupted and subsequently assimilated back to the chromosphere along paths either side of the prominence body, but due the relative faint intensity compared to the surrounding material, this motion is not distinguished well by the spatial RHT. The residual phase (Figure~\ref{fig:Fig9}(d)) shows similar structure as the plasma continues to drain, with the terminus of the dominant drainage curve moving to a lower latitude, compressing the arch structure and resulting in more strongly-radial angles.  

The angle distributions for each phase in Figure~\ref{fig:Fig9}(e) are consistent with both observational evidence and the key structural features revealed by the spatial RHT. We in  particular see a contrast between the first two phases, which show a range of angles, and the latter two phases, which are dominated by radial angles. Radial angles are due to the prominence column which remains, at least in part, throughout the event. The destabilisation in the swirl phase introduces more tangential components leading to a complex interplay between radial and tangential orientations, which gradual subsides as activity decreases in the later phases.

\section{Structural Classification of Prominences} \label{sec:Sect5}
We introduce a classification system based on the structural information provided by the spatial RHT which classifies prominences as either radial or tangential oriented. To construct this classification regime, we extract histograms of each timeseries showing the percentages of pixels for each angle, using bar widths of 5 degrees. We then weight the distribution for radial and tangential contributions, using weights ranging from 1 to 0, such that bins at $\pm 90$ degrees have a radial (tangential) contribution of 1 (0), and bins at 0 degrees have a radial (tangential) contribution 0 (1). Summing these two weighted distributions and taking the ratio provides us with the classification, as defined by:
\begin{equation}
\textrm{Classification} = 
\begin{cases*}
\textrm{Radial}, & if $\frac{R}{T} \geq 1$, \\
\textrm{Tangential}, & if $\frac{R}{T} < 1$,
\end{cases*}
\label{eq:Eq4}
\end{equation}
with $R$ and $T$ representing the weighted sums of the radial and tangential contributions, representatively. We define the `strength' of the classification by assessing the absolute difference of the ratio, $\frac{R}{T}$, from 1:
\begin{equation}
\textrm{Strength} = \left| \frac{R}{T} - 1 \right|~\textrm{.} \\
\label{eq:Eq5}
\end{equation}
The strength quantifies the degree of imbalance in the distribution toward a particular classification and takes values from 0 to 1, with greater values denoting a stronger classification.

\begin{figure}[h]
\graphicspath{ {./images/} }
    \centering
    \includegraphics[width=\linewidth]{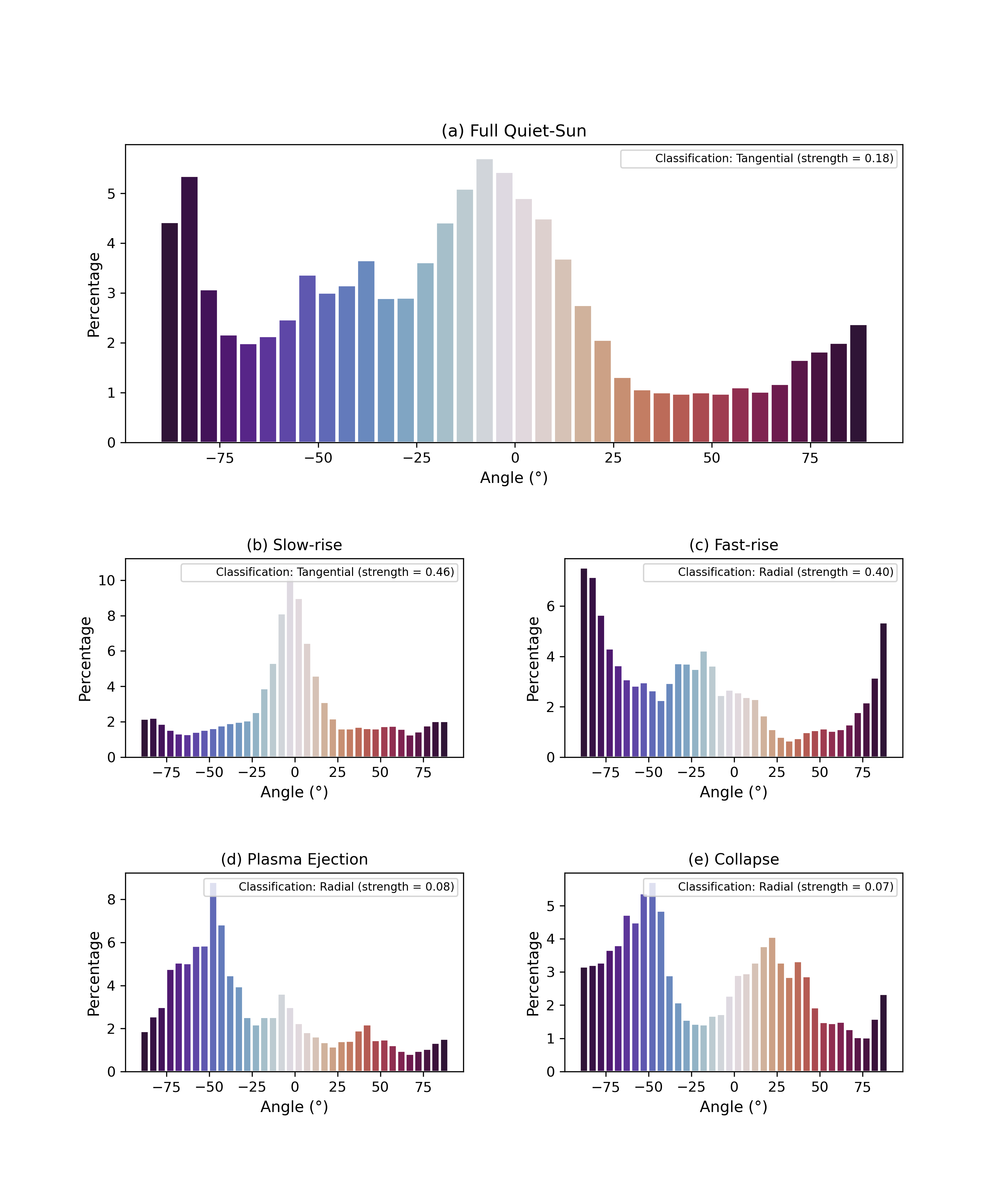} 
    \caption{Histograms for the percentage occupation of spatial RHT mean angles across different timeseries' in the quiet-Sun event, organised into 5 degree bins, showing: (a) full event, slow-rise (b), fast-rise (c), plasma ejection (d), and collapse (e).}
    \label{fig:Fig10}
\end{figure}

Figure~\ref{fig:Fig10} shows the histograms and classification for the quiet-Sun event, split into full (top) and  regional (middle and bottom) timeseries distributions. The percentage contributions of radial and tangential features for the event and the phases are shown in Table~\ref{tab:Tab3}. The full distribution (Figure~\ref{fig:Fig10}(a)) contains two dominant peaks, one near $-90$ degrees and one near 0 degrees, showing a mixture of strong radial and tangential features. Radial features are predominantly oriented clockwise (blue features in Figure~\ref{fig:Fig7}) and derive from the several pathways from which the prominence mass is siphoned back to the chromosphere during relaxation. These pathways also contribute to tangential activity, although the bulk of the tangential contribution is from the existed plasma on the limb during the long slow-rise phase (Figure~\ref{fig:Fig7}(a)), which constitutes almost 60\% of the entire event. The tangential contribution is $54.95\%$, around $10\%$ greater than the radial contribution. Consequently, it is clear why the overall classification here is tangential, with a strength of 0.18.

The regional information (Figure~\ref{fig:Fig10}(b) to (e)) provides more detail how the structure evolves with the activation of the prominence. The event begins with a significant contribution ($64.93\%$) of tangential angles (Figure~\ref{fig:Fig10}(b)), reflected in the high strength value for the classification. The fast-rise phase (Figure~\ref{fig:Fig10}(c)) shows a stark contrast with radial angles now dominating ($58.26\%$), with a strength similar in magnitude as previously. The classification strength rapidly decreases as the drainage lines permeate the event (Figure~\ref{fig:Fig10}(d)), denoted by the clockwise radial activity around -50 degrees. Anti-clockwise oriented features, seen at lower latitudes, occupy more of the collapse phase as more of the plasma is drained away (Figure~\ref{fig:Fig10}(e)). There is less than $4\%$ between the contributions in this phase. Although the final two phases display a greater range of feature orientation, they are still show more radial features ($51.82\%$ and $51.59\%$) and are classified as such. 

\begin{figure}[h]
\graphicspath{ {./images/} }
    \centering
    \includegraphics[width=\linewidth]{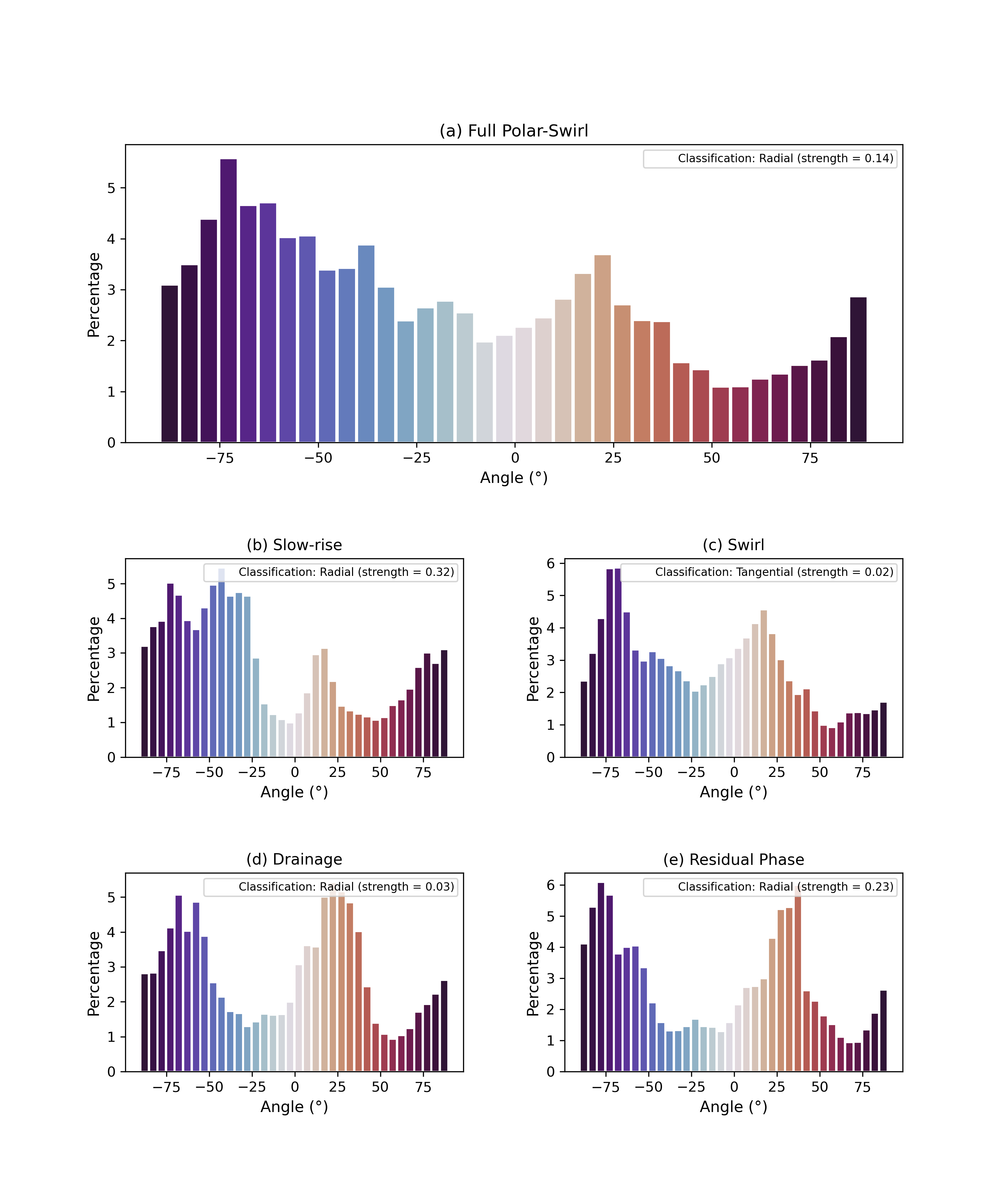} 
    \caption{Histograms for the percentage occupation of spatial RHT mean angles across different timeseries in the polar-swirl activation, organised into 5 degree bins, showing: (a) full event, slow-rise (b), swirl (c), drainage (d), and residual phase (e).}
    \label{fig:Fig11}
\end{figure}

Figure~\ref{fig:Fig11} shows the histograms and classification for the polar-swirl event, split into full (top) and  regional (middle and bottom) timeseries distributions. The percentage contributions of radial and tangential features for the event and the phases are shown in Table~\ref{tab:Tab4}. The full distribution (Figure~\ref{fig:Fig11}(a)) has radial classification with strength 0.14 and contribution of $53.18\%$, with most features oriented in a clockwise radial direction. These angles are primarily from the lower prominence body, which remains largely undisturbed throughout the activation process. The swirling motion introduces features with a more diverse range of angles, some of which lie close to the tangential as they curve from the prominence to higher latitude positions.

The regional information (Figure~\ref{fig:Fig11}(b) to (e)) indicates a broader variance of the structure. Figure~\ref{fig:Fig11}(b) shows a disproportional amount of radial angles ($56.96\%$) at the start of the event with a radial classification strength of 0.32, originating from the prominence column on the limb. A smaller tangential contribution can be seen around the 20 degree bin. This tangential activity increases as the event progresses, driven by the increasing amount of plasma transported laterally above the limb, and results in the classification alternating between radial and tangential (Figure~\ref{fig:Fig11}(c) and (d)). The low strength values signify an almost equal contribution of radial and tangential angles, with differences of around $1\%$ for the swirl and drainage phases. As the event relaxes in the residual phase, drainage field lines compress, providing stronger radial contribution ($55.18\%$) and classification (Figure~\ref{fig:Fig11}(e)).

\begin{table*}[bt]
\centering
\caption{Radial and Tangential angle contributions in the quiet-Sun eruption}
\label{tab:Tab3}
\begin{tabular}{c c c}
\hline
\hline
Phase & Radial Contribution (\%) & Tangential Contribution (\%) \\
\hline
Full & 45.05 & 54.95 \\
Slow-rise & 35.07 & 64.93 \\
Fast-rise & 58.26 & 41.74 \\
Plasma Ejection & 51.82 & 48.18 \\
Collapse & 51.59 & 48.41 \\
\hline
\end{tabular}
\end{table*}

\begin{table*}[bt]
\centering
\caption{Radial and Tangential angle contributions in the polar-swirl activation}
\label{tab:Tab4}
\begin{tabular}{c c c}
\hline
\hline
Phase & Radial Contribution (\%) & Tangential Contribution (\%) \\
\hline
Full & 53.18 & 46.82 \\
Slow-rise & 56.96 & 43.04 \\
Swirl & 49.41 & 50.59 \\
Drainage & 50.72 & 49.28 \\
Residual Phase & 55.18 & 44.82 \\
\hline
\end{tabular}
\end{table*}

\section{Assessment of the Spatial RHT Algorithm} \label{sec:Sect6}
The RHT algorithm is a versatile method of obtaining structural detail of curvilinear features within a image. We have shown here a breakdown of the key steps in the spatial RHT procedure, along with its adaptation to solar prominence structural and evolutional analysis. Additionally, the change of reference into the polar frame introduced in Section~\ref{sec:Sect4} enables a more intuitive handling of the data and subsequent interpretation of structural alignment. Nevertheless, as with any computational algorithm, there are many factors that may impede or improve the results and overall performance. Knowledge of these factors is key to optimising the spatial RHT output. We list these factors below with suggestions on how to improve them.

\paragraph{\textbf{Parameters}} 
The first obvious influence on the spatial RHT is the choice of parameters. Table~\ref{tab:Tab5} shows a description of the spatial RHT tuneable parameters, along with the values used for this study, and notes on choosing appropriate values. It is also possible to tune parameters that exist within the main RHT code to create more stringent detection criteria. One could achieve this by increasing values of the adaptive threshold value, $f$, or the condition $\bar{R} \geq 0.75$. Such alterations, however, are largely unnecessary, as these parameters are already optimized within the code and have consistently demonstrated good performance. Most effort should be spent on tailoring the tuneable parameters to the desired dataset and detected features.

\begin{table*}[hbt]
\centering
\caption{Spatial RHT parameters.}
\label{tab:Tab5}
\begin{tabular}{l p{4cm} >{\centering\arraybackslash}p{2cm} p{5cm}}
\hline
\hline
\colhead{Parameter} & \colhead{Description} & \colhead{Value Used} & \colhead{Notes} \\
\hline
\texttt{smr\_xy} (spatial filter) & Controls the spatial smoothing of features. & 30 & Increase for greater filtering and feature width at the expense of information. \\
\texttt{med\_xy} (median filter) & Applies median filtering to smooth data. & 3 & Increase for stronger filtering effect but may lose fine details. \\
\texttt{wlen} (kernel size) & Determines the kernel width for feature detection. & 51 & Must be an odd-numbered integer; tailor to the width of detected features. \\
\texttt{nlev} (noise level) & Adjusts for the background noise level. & 10 & Any positive value; tailor to intensity level of background noise. \\
\hline
\end{tabular}
\end{table*}

\paragraph{\textbf{Filtering}}
The RHT algorithm is heavily dependent on the quality of the filtering and the ability to isolate the curvilinear features within the image. As mentioned above, the filtering is controlled by the \texttt{smr\_xy} and \texttt{med\_xy} parameters, but can also be additionally applied to images as a preprocessing step. The quality of filtering for detection depends on the ratio of the feature pixel intensity with the background pixel intensity, as this creates a strong definition of an `edge' along which the RHT will interpret as a curvilinear feature(this in turn comes with the caveat of image edges being detected as linear features which have to be manually removed). Low intensity or faint features will avoid detection unless care is taken to artificially enhance them against the background intensity, although this can often lead to an enhancement of noise in the image. This problem was identified by \citet{2022ApJ...931L..27S}, where the authors reported challenges in coronal rain detection in cases of low emission. For Solar image data, better results may be achieved by applying filtering techniques on the raw data such as the radial gradient filter \citep{2006SoPh..236..263M}, Multi-Gaussian Normalisation \citep{2014SoPh..289.2945M}, or the WOW technique \citep{2023A&A...670A..66A}, to name a few.

\paragraph{\textbf{Input Image Quality}}
A perhaps obvious factor, is the quality of the input image, in the absence of any filtering. For solar physics applications, this is largely determined by the choice of instrument (spatial resolution) and wavelength channel used, which are fixed and not variables that can be modified or improved like the filtering parameters. Isolating prominences in the 304 \AA\ channel is often challenging due to the hotter Si \textsc{xi} 303.3 Å\ emission line which obscures much of the finer structural detail and is especially prevalent in off-limb observations \citep{2000SoPh..195...45T, 2010A&A...521A..21O}. Some degree of noise is filtered through the \texttt{nlev} parameter, although setting this too high will remove pixels of interest. Clearer results can be obtained though other channels, for example the 171 \AA\ channel, although is more suited for coronal loops and arcades rather than prominences. This aspect highlights the importance of filtering, especially if there are observational nuances within the wavelength channel for the dataset.

\paragraph{\textbf{Image features}}
It is important to consider the features within the dataset images, not just as they determine the value of the \texttt{wlen} parameter, but they also provide an indication of how well the RHT algorithm will perform. The RHT is optimised for curvilinear features, particularly those which are well isolated from any background noise (refer to panel (d) of Figure~\ref{fig:Fig5}). Prominences themselves are not strict curvilinear structures, and may be observed in a variety of amorphous geometries, but are generally isolated against the dark background in off-limb observations\footnote{We note that, whilst stronger curvilinear characteristic are seen in observations of filaments on-disk, further filtering would likely be required to isolate them in H$\alpha$ images.}. They can also exhibit fine-scale linear features as has been observed in the vertical threads of hedgerow prominences \citep{1976SoPh...49..283E,2008ApJ...676L..89B, 2008ApJ...689L..73C}. The spatial RHT excels in situations where there is clear fine structure, but struggles with bulk mass. The same can be said of the spatial RHT outputs, which provides fine-scale structural information rather than bulk-wise properties. We see this  expressed in our results. The plasma ejection and collapse phase of Figure~\ref{fig:Fig1}(c) and (d) show clear curvilinear features in the drainage paths which the spatial RHT is able to isolate effectively (Figure~\ref{fig:Fig9}(c) and (d). When the structural detail is more obscure, such as the Swirl phase seen in Figure~\ref{fig:Fig2}(b), the spatial RHT struggles to identify curvilinear features, resulting in more sparsely connected structures (Figure~\ref{fig:Fig9}(b)). 

\begin{figure}[h!]
\graphicspath{ {./images/} }
    \centering
    \includegraphics[width=\linewidth]{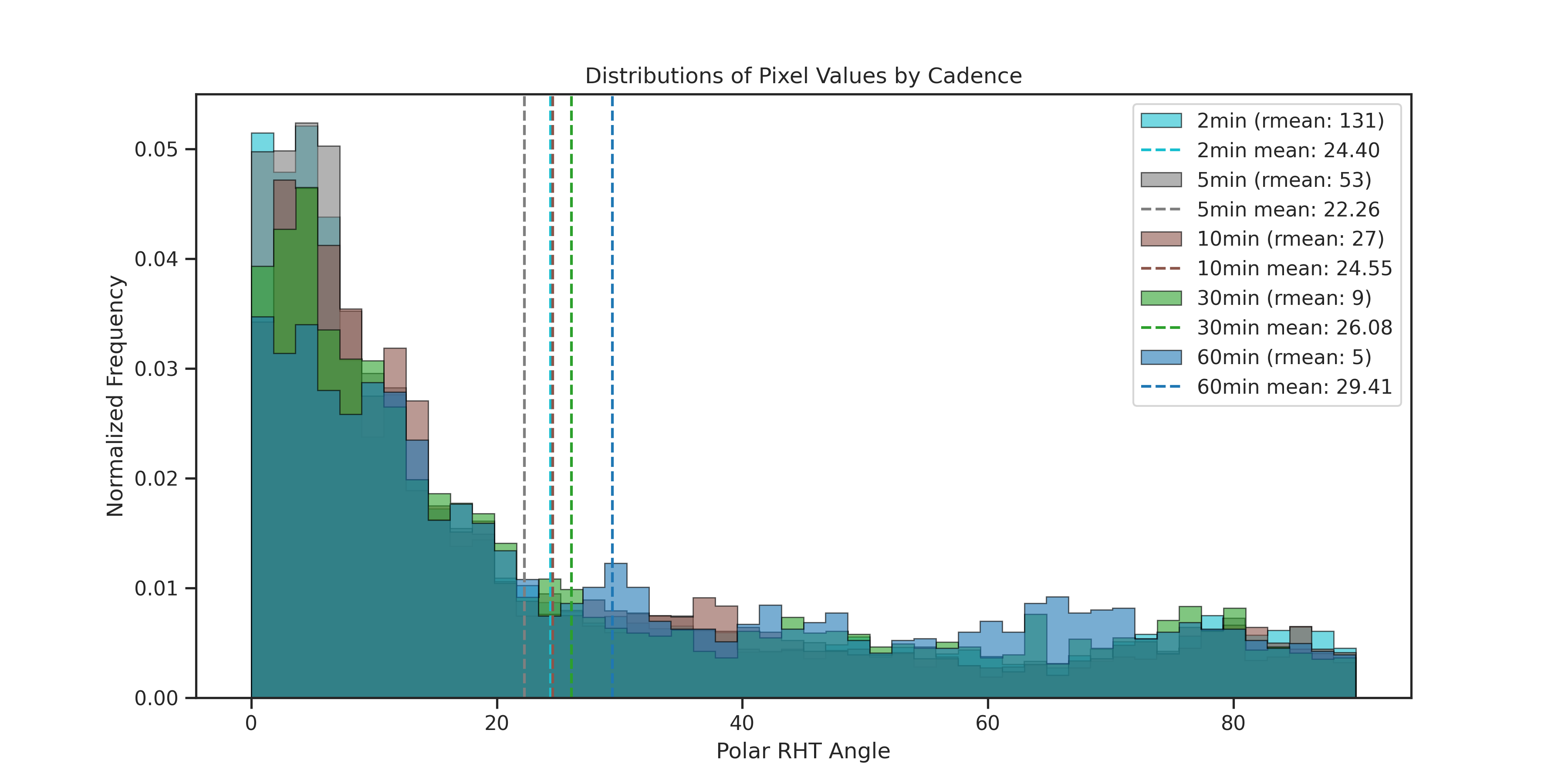} 
    \caption{Histogram of detected pixels for the first image of the quiet-Sun eruption, taken at time 14:00:05.13 UT, using different cadence values. The cadence value determines the temporal dimension of the data and hence the value of temporal filter kernel. Accuracy of the assigned spatial RHT angles increases with smaller cadence values.}
    \label{fig:Fig12}
\end{figure}

\paragraph{\textbf{Timeseries}} 
Although the spatial RHT is restricted to just the image spatial dimensions, the filtering process contains a temporal smoothing filter where the kernel is equal to the number of images within the timeseries (the temporal dimension). This is a method of calibrating images in a timeseries, accounting for any large changes in the dynamic range (for example due to flares). Results will therefore differ depending on the length of the input timeseries, which in our study is determined by the cadence of observations. Figure~\ref{fig:Fig12} shows how the number of detected pixels varies with cadence, along with the different angle distributions. We use the spatial RHT result the first array in the quiet-Sun timeseries, taken at time 14:00 UT.  

This test conveys the inverse relationship between the choice of cadence and the accuracy of the spatial RHT algorithm. Larger values of the cadence have smaller temporal dimensions (less images in the dataset), which in turn leads to a larger temporal kernel. This can lead to over-smoothing of the data, whereby structural information may be lost, thus decreasing the accuracy of the assigned spatial RHT angles. Likewise, smaller cadences have a larger temporal dimension (more images in the dataset), and thus a smaller temporal kernel - more data is preserved after smoothing. This emphasises the importance of the cadence on the spatial RHT results even if the dataset covers the same temporal domain. \\

It is important here to remind that all quantities calculated by the spatial RHT are in the plane-of sky. Projection effects can distort the true orientation of structures, potentially resulting in misleading interpretations of the angles calculated by the spatial RHT. Verification of the true motion can be carried out by considering further line-of-sight measurements, multi-perspective observations from other instruments, or through 3D reconstructions \citep{2010A&A...514A..68S,2017A&A...606A..30S}. Such methods are beyond the scope of our study as we focus only on single-instrument analysis, but must be considered if further analysis into prominence motion is required.

\section{Discussion} \label{sec:Sect7}

In our study we endeavour to address two questions:
\begin{itemize}
    \item How can the spatial RHT routine aid in visualisation of prominence structure and localised dynamics of prominence eruptions/activation? In other words, can the spatial RHT be successfully applied to prominence dynamics and return effective results?
    \item Can we utilise the structural information provided by the RHT to construct a classification regime whereby prominence events are classed based on the feature orientation? Through this, we aim to class prominence events as either tangential or radial, along with a measure of the contributions of radially/tangentially aligned features. 
\end{itemize} 

Application of the spatial RHT for structural information has been applied by \citet{2017SoPh..292..132S}, who showed how the algorithm can distinguish both the orientations of EUV coronal loops and the fine chromospheric fibril structure of an active region using TRACE and IBIS data, respectively. The author, alongside \citet{ 2022ApJ...931L..27S, 2023ApJ...950..171S}, has also applied the RHT to the semi-automatic detection of coronal rain. In these studies the authors are able to obtain structural properties of the rain showers, namely the length and width, track rain trajectories, and determine projected speeds of the rain\footnote{It should be noted that these studies utilised temporal RHT, which extends the filtering and RHT routine to account for temporal changes along a trajectory, in addition to the spatial RHT.}. In all cases, the routine is carefully tailored to suit the desired application. Our version of the spatial RHT is optimised for prominence detection and thus differs with the handling of certain parameters, filtering, etc.. The results from Figures~\ref{fig:Fig7} and~\ref{fig:Fig9}, and the comparisons with the observational SDO/AIA data, indicate that our choice of parameters is sensible. For both events studied, the spatial RHT is able to isolate the key structural features of the prominence, alongside the trajectories of the plasma. We particular achieve noteworthy results in cases where the pathways (field lines) are strongly represented, as in the drainage pathways in both events (panels (B) and (c) of Figures~\ref{fig:Fig7} and~\ref{fig:Fig9}). We are able to discern prominence evolution after splitting the events into distinct phases and analysing the distributions of angles detected by the spatial RHT. Peaks in the distribution show the most dominant angle position. During the activation of the events, we see dramatic changes in the distributions corresponding to the dynamics of the event.

The spatial RHT operates over image coordinates, therefore, for appropriate consideration of the Solar reference frame, we transform into polar coordinates. Radial (parallel to the solar limb) and tangential (perpendicular to the solar limb) features in our polar reference frame are denoted by angles of $\pm90$ and 0 degrees, respectively. Furthermore, the definition of the positive/negative angles for radial structures as anticlockwise/clockwise orientated is consistent with the motion from the movies. Classification of the feature orientation is achieved through weighting the polar angle distributions, from which we are able to obtain a strength value of the classification, along with the contribution of radial and tangential features. The histograms and resulting radial/tangential contributions show greater significance due to the large number of points in the distributions, providing a more accurate picture than spatial RHT averages from Figures~\ref{fig:Fig7} and~\ref{fig:Fig9}. They also provide greater insight into the dynamics of the events, particular during each individual phase (see tables~\ref{tab:Tab3} and~\ref{tab:Tab4}). 

\paragraph{\textbf{Quiet-Sun eruption}}
The quiet-Sun eruption is classified as tangential with a contribution of $54.95\%$ when taking results of the full event timescale. More detail is provided in the regional classification, which shows strong tangential characteristics ($64.93\%$) in the precursor phase of the eruption, which then shifts to radial alignment ($58.26\%$) with the onset of the eruption. We interpret this as the tangentially oriented prominence material accumulating on the limb during the slow-rise phase, which gradually rises and erupts along the radial direction, thus explaining the of interchange of radial and tangential alignment. Tangentially aligned features begin to return near the end of the eruption, though as plasma drainage pathways along the field lines down into the chromosphere. From the observational data, all of the ejected material appears to be involved in this drainage process, indicating that the ejection velocity was not sufficient enough for it to escape out of the solar atmosphere. The ejected material is too faint to be detected by the spatial RHT, but it clearly reveals the drainage paths, which seem to return to the original eruption site, possibly back to the filament channel to re-form (the emergence of plasma in the lower portion of the eruption in panels (c) and (d) from Figure~\ref{fig:Fig7} may suggest this). This would evidence a partial or failed eruption. It is known that the kink instability plays a role in partial/failed eruptions and subsequent filament re-formation \citep{2007SoPh..245..287G}, although further analysis is needed to confirm if this is the case here (projection effects make direct observation of filament kinking difficult). It is also possible that not all of the prominence material takes part in the eruption. Indeed, there remains some tangential features within the lower portion of the prominence seen in panel (b) of Figure~\ref{fig:Fig7}, which may indicate that not all of the magnetic structure undergoes eruption.

\paragraph{\textbf{Polar-swirl activation}}
The polar-swirl activation is classified as radial with a contribution of $53.18\%$ for the full event. Investigating the individual phases, however, shows a large interplay of radial and tangential feature orientation. The initial activation of the polar crown prominence contrasts the quiet-Sun and is dominated by radial features $56.96\%$, beginning with a largely radially-aligned column of plasma from the foot-point on the East limb. Of these radial features, the majority are oriented clockwise, that is, pointing towards the solar north pole, indicating interactions of the prominence with a region situated at a higher latitude. As the activation of the prominence progresses the system destabilises, introducing more tangential contributions ($50.59\%$) due to lateral plasma flows from the prominence to this higher latitude point. Despite the top part of the prominence column breaking during the activation, the lower portion of the column persists throughout the remainder of the event, providing a strong radial contribution to balance with the tangential motion. Consequently, we see little difference in the contributions in the swirl and drainage phases of the activation. Radial features start to dominate again ($55.18\%$) in the residual phase where the swirling motion has largely dissipated and drainage paths become more compressed. \\

Our classification system shares similarity with that of \citet{2000ApJ...537..503G} and \citet{2003ApJ...586..562G}, where the orientation of motion is correlated with the type of prominence, with radial motion corresponding to an eruptive prominence, and tangential motion to a active prominence. Although the spatial RHT process does not discern the motion, we can still apply the same logic here by examining the phases where the prominence is activated. The quiet-Sun eruption, as the name implies, can be classified as an eruptive prominence, as some of the ejected material appears to escape the solar gravitational field. The RHT reveals predominantly radial structure during the eruption processes in the fast-rise and plasma ejection phases (Figure~\ref{fig:Fig10}(c) and (d)). Likewise, the polar-swirl activation demonstrates characteristics of an active prominence, as material does not appear to escape the solar gravitational field, and instead is transported to another point on the solar surface. In the swirl phase the spatial RHT detects a large tangential component to the structure, activity which persist throughout the remainder of the event (see Figure~\ref{fig:Fig11}(c) to (e)). It should be noted that whilst the regional contributions convey detailed information on the event dynamics, the full contributions do not adequately capture these more nuanced dynamics leading to a discrepancy in the overall classification with the individual breakdowns.  

\section{Conclusion} \label{sec:Sect8}
Our study has shown the application of the spatial RHT algorithm to two distinct prominence dynamics events, the quiet-Sun prominence eruption, and the polar-swirl prominence activation. In particular we have demonstrated the successful adaptation of the code to produce a classification output based on the structural orientation of fine-scale features in the plane-of-sky, denoting events as either tangential or radial. We obtain results for the full timeseries of both events as well as different phases within, where the dynamics are more consistent. The information uncovered by the spatial RHT results underscores the variability in the plasma dynamics for both events, driven by the distinct physical conditions and processes which underpin their evolution, and is additionally consistent with the observational data. This study serves as a proof-of-concept for the application of the spatial RHT towards prominence dynamics. \\

The spatial RHT algorithm can be extended into the temporal domain (temporal RHT), which would allow for estimates of the velocities of plasma eruption and drainage, allowing for more detailed analysis of prominence eruption/activation events. Incorporating velocity calculations will then form part of the classification scheme for eruptive prominences. The classification of prominences is important for discriminating between the different models of eruptions and to provide constraints on the dynamo action over the solar cycle.\\

\section*{\textbf{Acknowledgments}}
All authors acknowledge the UK Research and Innovation (UKRI) Science and Technology Facilities Council (STFC) for support from grant No. ST/W006790/1. This study has been supported by the STFC Centre for Doctoral Training in Data Intensive Science (NUdata), as a collaboration between Northumbria and Newcastle Universities. The authors would like to thank T. Schad for help with the RHT code and S. Şahin for providing the IDL version of the code, which was used as a template for our python code. SDO/AIA data are courtesy of NASA/SDO and the AIA science team, and were accessed via the Joint Science Operations Center (JSOC) at Stanford University. The authors are willing to provide the data and code upon reasonable request. We are grateful to the referee for their insightful and informative comments on the manuscript.

\bibliography{references,references2}
\bibliographystyle{aasjournal}

\end{document}